\documentclass[11pt]{article}
\usepackage{fullpage,citesort,epsfig,graphics,amsbsy}%,newcommand1}
\usepackage{psfrag}
\usepackage[small]{caption}
\newcommand{\beq}{\begin{equation}}
\newcommand{\eeq}{\end{equation}}
\newcommand{\bea}{\begin{eqnarray}}
\newcommand{\eea}{\end{eqnarray}}

\newcommand{\be}{\begin{equation}}
\newcommand{\ee}{\end{equation}}
\newcommand{\bq}{\begin{eqnarray}}
\newcommand{\eq}{\end{eqnarray}}

\def\math{\mathsurround=0pt }
\def\leftrightarrowfill{$\math \mathord\gets \mkern-6mu \cleaders\hbox{$\mkern-2mu \mathord- \mkern-2mu$}\hfill
 \mkern-6mu \mathord\to$}
\def\overleftrightarrow#1{\vbox{\ialign{##\crcr
     \leftrightarrowfill\crcr\noalign{\kern-1pt\nointerlineskip}
     $\hfil\displaystyle{#1}\hfil$\crcr}}}

\newcommand{\bfs}{\boldsymbol}
\let\l=\lambda

\def\nn{\nonumber} \def\bd{\begin{document}} \def\ed{\end{document}}
\def\ds{\documentstyle} \let\fr=\frac \let\bl=\bigl \let\br=\bigr
\let\Br=\Bigr \let\Bl=\Bigl
\let\bm=\bibitem
\let\na=\nabla
\let\pa=\partial \let\ov=\overline
\def\ft#1#2{{\textstyle{{\scriptstyle #1}\over {\scriptstyle #2}}}}
\def\fft#1#2{{#1 \over #2}}
\def\vp{\varphi}
\def\sst#1{{\scriptscriptstyle #1}}
\def\oneone{\rlap 1\mkern4mu{\rm l}}
\def\td{\tilde}
\def\wtd{\widetilde}
\def\dalemb#1#2{{\vbox{\hrule height .#2pt
        \hbox{\vrule width.#2pt height#1pt \kern#1pt
                \vrule width.#2pt}
        \hrule height.#2pt}}}
\def\square{\mathord{\dalemb{6.8}{7}\hbox{\hskip1pt}}}
\def\wtd{\widetilde}
\def\R{\rlap{\rm I}\mkern3mu{\rm R}}
\def\im{{\rm i}}
\def\tilg{\tilde{g}}
\def\tilF{\tilde{F}}
\def\tilA{\tilde{A}}
\def\varf{\varphi}
\def\tilf{\tilde{\phi}}
\def\tilh{\tilde{h}}
\def\rme{{\rm e}}
\def\ep{\epsilon}
\def\0{{(0)}}
\def\9{{(9)}}
\def\8{{(8)}}
\def\7{{(7)}}
\def\6{{(6)}}
\def\5{{(5)}}
\def\4{{(4)}}
\def\3{{(3)}}
\def\2{{(2)}}
\def\1{{(1)}}
\newcommand{\trace}{{\rm Tr}}
\newcommand{\ub}{\overline{U}}
\newcommand{\vb}{\overline{V}}
\newcommand{\uh}{\widehat{U}}
\newcommand{\vh}{\widehat{V}}
\newcommand{\ubh}{\overline{\widehat{U}}}
\newcommand{\vbh}{\overline{\widehat{V}}}
\newcommand{\lb}{\bar{\l}}
\newcommand{\Fb}{\overline{F}}
\newcommand{\Fh}{\widehat{F}}
\newcommand{\Fbh}{\overline{\widehat{F}}}
\newcommand{\Ab}{\overline{A}}
\newcommand{\Ah}{\widehat{A}}
\newcommand{\Abh}{\overline{\widehat{A}}}
\newcommand{\Gb}{\overline{G}}
\newcommand{\Gh}{\widehat{G}}
\newcommand{\Gbh}{\overline{\widehat{G}}}
\newcommand{\Pb}{\overline{P}}
\newcommand{\Ph}{\widehat{P}}
\newcommand{\Pbh}{\overline{\widehat{P}}}
\newcommand{\Qb}{\overline{Q}}
\newcommand{\Qh}{\widehat{Q}}
\newcommand{\Qbh}{\overline{\widehat{Q}}}
\newcommand{\Bb}{\overline{B}}
\newcommand{\Bh}{\widehat{B}}
\newcommand{\Bbh}{\overline{\widehat{B}}}
\newcommand{\fhns}{\hat{F}^{\rm (NS)}}
\newcommand{\fhrr}{\hat{F}^{\rm (RR)}}
\newcommand{\ahns}{\hat{A}^{\rm (NS)}}
\newcommand{\ahrr}{\hat{A}^{\rm (RR)}}
\newcommand{\hhrr}{\hat{H}^{\rm (RR)}}
\newcommand{\hchi}{\hat{\chi}}
\newcommand{\hphi}{\hat{\phi}}
\newcommand{\htau}{\hat{\tau}}
\newcommand{\cG}{{\cal G}}
\newcommand{\cGb}{\overline{{\cal G}}}
\newcommand{\cH}{{\cal H}}
\newcommand{\cP}{{\cal P}}
\newcommand{\cPb}{\overline{{\cal P}}}
\newcommand{\cQ}{{\cal Q}}
\newcommand{\cQb}{\overline{{\cal Q}}}
\newcommand{\cM}{{\cal M}}
\newcommand{\cN}{{\cal N}}
\newcommand{\cO}{{\cal O}}
\newcommand{\cD}{{\cal D}}
\newcommand{\cL}{{\cal L}}
\newcommand{\cA}{{\cal A}}
\newcommand{\cB}{{\cal B}}
\newcommand{\hg}{\hat{g}}
\newcommand{\cE}{{\cal E}}
\newcommand{\vpp}{\mbox{$\langle{\scriptstyle++}\rangle$}}
\newcommand{\vmp}{\mbox{$\langle{\scriptstyle-+}\rangle$}}
\newcommand{\vppp}{\mbox{$\langle{\scriptstyle+++}\rangle$}}
\newcommand{\vmpp}{\mbox{$\langle{\scriptstyle-++}\rangle$}}
\newcommand{\vpmp}{\mbox{$\langle{\scriptstyle+-+}\rangle$}}
\newcommand {\Kftw} {K^\land_{43}}
\newcommand {\Kttw} {K^\land_{32}}
\newcommand {\Ktow} {K^\land_{21}}
\newcommand {\Kofw} {K^\land_{14}}
\newcommand {\Kftv} {K^\lor_{43}}
\newcommand {\Kttv} {K^\lor_{32}}
\newcommand {\Ktov} {K^\lor_{21}}
\newcommand {\Kofv} {K^\lor_{14}}
\newcommand {\Po} {k^+_1}
\newcommand {\Pt} {k^+_0}
\newcommand {\Pth} {k^+_3}
\newcommand {\Pf} {k^+_2}
\newcommand {\Kp} {q^+}
\newcommand {\T} {T_1}
\newcommand {\Uu} {T_0}
\newcommand {\eS} {T_3}
\newcommand {\V} {T_2}
\newcommand {\invt} {(-t)}
\newcommand {\pref} {\frac{1}{8\pi^2}}
\newcommand {\rleft} {\Pt<\Kp<\Pth}
\newcommand {\rmid} {\Pth<\Kp<\Po}
\newcommand {\rright} {\Po<\Kp<\Pf}
\newcommand {\Anei} {A_{\land\land\lor\lor}}
\newcommand {\Aalt} {A_{\land\lor\land\lor}}
\newcommand {\Nc} {}

\begin{document}
%\draft

\setlength{\captionmargin}{20pt}
\begin{titlepage}
\begin{flushright}
\phantom{UFIFT-HEP-10-}
\end{flushright}

\vskip 2.5cm

\begin{center}
\begin{Large}
{\bf Resolution of Infrared Divergences in
Gluon-Gluon Scattering Regulated on a 
Lightcone Worldsheet Lattice\footnote{Supported 
in part by the Department
of Energy under Grant No. DE-FG02-97ER-41029.}}
\end{Large}

\vskip 2cm
{\large 
Charles B. Thorn\footnote{E-mail  address: {\tt thorn@phys.ufl.edu}}
}
\vskip0.20cm
{\it Institute for Fundamental Theory\\
Department of Physics, University of Florida,
Gainesville FL 32611}
%\vskip12pt(\today)
\vskip 1.0cm
\end{center}

\begin{abstract}\noindent
We improve and update the 
discussion, given some years ago 
by my collaborators and me, of infrared divergences and Bremsstrahlung 
in one-loop gluon scattering
probabilities in lightcone gauge. In that work, we
showed that adding soft and collinear gluon radiation,
satisfying simple Lorentz invariant constraints, not 
only cancelled all IR divergences, but resulted in
compact expressions for the consequent scattering probabilities.
Here we impose less restrictive (albeit noncovariant) constraints on
the unobserved radiation, which increases the high energy ($s$) fixed momentum
transfer ($t$) behavior of the total probabilities
from $-\ln^2s$ to $\ln s \ln t$, a behavior shared by
the (IR divergent) elastic probabilities. 
Using this new treatment we also make a much more
detailed comparison of the lightcone results to
covariant calculations using dimensional regularization,
finding complete agreement between the two styles of calculation.
\end{abstract}
\vfill
\end{titlepage}
\section{Introduction}
In this note we seek to clarify some aspects of 
the one loop QCD corrections to the scattering of glue by glue
in lightcone gauge as
calculated in \cite{chakrabartiqt,chakrabartiqt2}.
Those calculations employed an infrared
regulator motivated by the lightcone worldsheet lattice
\cite{gilest,bardakcit,thornsheet}, 
which discretizes both $p^+=(p^0+p^3)/\sqrt{2}$
and $ix^+=i(x^0+x^3)/\sqrt{2}$. In the field theory context,
the discretization of $x^+$ was immediately removed
after adoption of a worldsheet friendly ultraviolet cutoff $\delta$ on the
transverse momenta. But the discretization of $p^+=Mm$, $M=1,2,\cdots$ 
was retained as an infrared cutoff.

We first address some issues stemming from the novel manner in which 
soft and collinear gluon emission was included to resolve the
infrared divergences. Given the lightcone gauge setup, it
was natural to define a jet of momentum $P_i=k+p_i$, containing
two gluons, by the restriction on their momenta $k, p_i$:
\bea
{(p_i^+{\bfs k}-k^+{\bfs p}_i)^2\over k^+p_i^+}<\Delta^2,\qquad{\rm or}
\qquad \left({\bfs k}-{k^+\over p_t^+}{\bfs p}_i\right)^2 
< {k^+\over p_i^+}\Delta^2\; .
\label{jetdef}
\eea
In fact the left side of the first inequality is just $-(k+p_i)^2=-P_i^2$,
the invariant mass squared of the jet. The restriction simply
limits the ``virtuality'' of the jet compared to an on mass shell
gluon: this jet definition is Lorentz invariant. 
The interpretation of gluons as jets is of course part
of what is needed to define an infrared-safe scattering
probability. But one needs to include soft gluon
Bremsstrahlung as well. Usually this additional radiation
is defined by a condition such as $|k^0|<\epsilon$
on the energy of the extra gluon. Of course at the same time one
has to exclude such soft gluons from the jet definition, to
avoid double counting. But in \cite{chakrabartiqt2}, 
we pursued a Lorentz covariant alternative to this, namely:
include in the Bremsstrahlung
part of the calculation any extra gluon with momentum $k^\mu$ satisfying
the jet condition (\ref{jetdef}) for at least one of the external
legs of the core process. Indeed, we showed that adding just this
real radiation to the one loop-contributions to the
elastic scattering probabilities cancelled all the infrared divergences,
in accord with the Lee-Nauenberg theorem \cite{leenauenberg}.
Moreover, this treatment produces nice compact 
Lorentz invariant expressions for the
total scattering probabilities [see (\ref{jetprobfinal})
and (\ref{jetprobfinal2}) in Section 3]. 

The only problem is that these formulas
have an awkward high energy (Regge) limit $s\to\infty$ with $t$ fixed.
Inspection of the formulas shows that the dominant behavior in this limit
goes as $-\ln^2 s$, with a {\it negative} coefficient! This clashes with
the absence of a $\ln^2 s$ behavior in the known covariant dimensionally
regulated elastic scattering amplitude \cite{kunsztst}: the double log
terms in these elastic amplitudes are of the form $\ln s \ln t$.
The absence of $\ln^2 s$
behavior at one loop is compatible with the hypothesis that the
higher order corrections put the gluon
on a Regge trajectory of order $g^2$: a behavior $s^{1+g^2f(t)}
\to s(1+g^2f(t)\ln s)$. Indeed, recent
interest in this possibility \cite{schnitzer} in connection with the AdS/CFT
correspondence was one motivation for the present update.  
Since the $-\ln^2 s$ behavior found in \cite{chakrabartiqt2}
includes only part of the
inelastic Bremsstrahlung processes, it is possible that including more
Bremsstrahlung will cancel this negative term. Had the $\ln^2 s$ term
been positive, it would have been bad news for Regge behavior!

We resolve this puzzle of too little
Bremsstrahlung in Section 3 by employing a more traditional,
and less restrictive, definition of 
soft gluon radiation: simply limit the $k^+$ of the extra gluon
$k^+<\kappa$. (To ensure that all components of $k$ are
soft, we also need to limit $k^-<O(\kappa)$. But, as we shall see,
the form of the
soft gluon emission amplitude suppresses $|{\bfs k}|>k^+$ sufficiently
to automatically satisfy this second restriction).
The consequent probabilities will
not be Lorentz invariant, but we show that in the center of
mass frame the $\ln^2 s$ behavior cancels as $s\to\infty$
with $t$ fixed. The calculations in this section differ from
those in \cite{chakrabartiqt2} in that they are done in
arbitrary transverse dimension.  But as long as $d$ is set to 2 {\it before}
integration over $k^+$, the two calculations are completely equivalent.

The second issue that we address involves the comparison of the results
of \cite{chakrabartiqt2} to previously known covariant 
calculations using dimensional regularization. In
\cite{chakrabartiqt2} we noted that the ratio of amplitudes
describing different gluon polarizations, from which
IR divergences cancel, agreed with
the known results \cite{kunsztst}.
However, the comparison of amplitudes with a given
polarization was obscured by the vagaries of IR divergences.
We must compare the results for total probabilities, since it is not
meaningful to compare the (gauge dependent) elastic amplitudes. 
So in section 4 we redo the
Bremsstrahlung calculations of Section 3 using dimensional
regularization throughout: all momentum integrals are
done {\it before} taking $d\to2$. We then combine these with the 
previously known
elastic scattering probabilities, obtained covariantly using dimensional
regularization, to obtain expressions for the
the total probabilities. These agree in every detail with the
results of Section 3. In this way we provide a definitive confirmation that
the discrete $p^+$ regularization, motivated by the
lightcone worldsheet lattice, provides a reliable treatment
of infrared divergences.

As in \cite{chakrabartiqt,chakrabartiqt2},
we organize the Feynman diagrams of the $SU(N_c)$ Yang-Mills theory
according to 't Hooft's large $N_c$ expansion \cite{thooftlargen}, 
and we calculate the one-loop planar diagrams surviving the $N_c\to\infty$
limit. The 't Hooft limit suppresses diagrams with quark loops, so 
they are not included here. We begin our discussion with a short
Section 2, which summarizes the Feynman rules for planar Yang-Mills
and sets our notation and conventions.
\section{Lightcone Feynman Rules for $N_c\to\infty$ Yang-Mills}
Here, we use the notation and conventions in Ref.~\cite{beringrt},
according to which the values of the 
non-vanishing three transverse gluon vertices are:
\bea
{{}\atop\mbox{\includegraphics[width=1.2cm]
{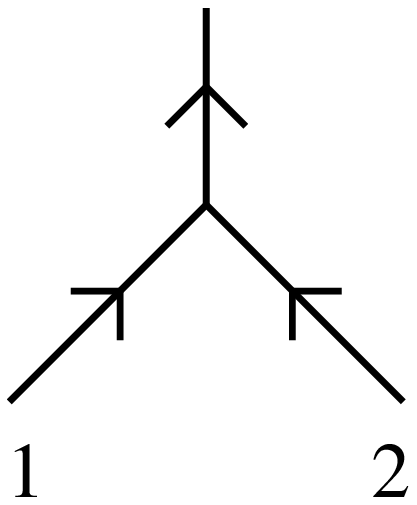}}}
\displaystyle\quad&=&{2gp_3^+\over p_1^+p_2^+}
\left(p_1^+{p_2^\land}-p_2^+{p_1^\land}\right)
={2gp_3^+\over p_1^+p_2^+}K^\land_{12}
\label{upupdown}\\
{{}\atop\mbox{\includegraphics[width=1.2cm]
{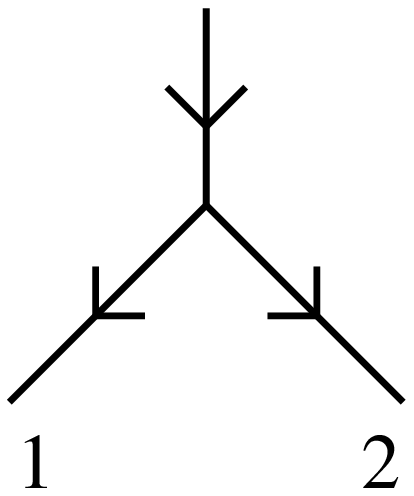}}}
\displaystyle\quad&=&{2gp_3^+\over p_1^+p_2^+}
\left(p_1^+{p_2^\lor}-p_2^+{p_1^\lor}\right)
={2gp_3^+\over p_1^+p_2^+}K^\lor_{12}
\label{downdownup}
\eea 
The quartic vertices in this helicity basis are given by
\bea
{{}\atop\mbox{\includegraphics[width=1.2cm]
{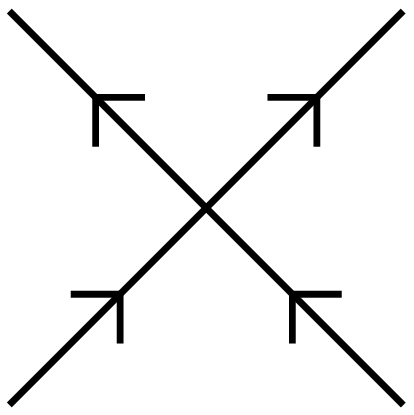}}}
\displaystyle\quad&=&-2{{g^2}{p^+_1p^+_3+p^+_2p^+_4\over
(p^+_1+p^+_4)^2}
\label{upupdowndown}}\\
{{}\atop\mbox{\includegraphics[width=1.2cm]
{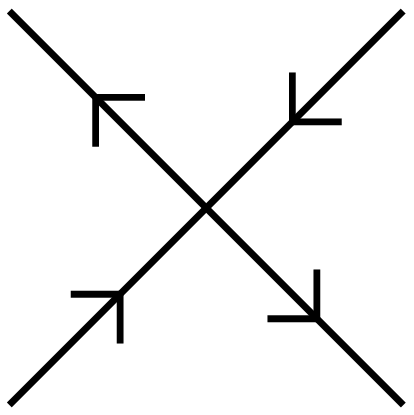}}}
\displaystyle\quad&=&+2{{g^2}\left({p^+_1p^+_2+p^+_3p_4^+\over
(p^+_1+p^+_4)^2}+{p^+_1p^+_4+p_2^+p^+_3\over
(p^+_1+p^+_2)^2}\right)}
\label{updownupdown}
\eea
In these expressions, $\land$ and $\lor$ label the $\pm$ helicity
of the gluon,
and $p_k^\land=(p_k^x+ip_k^y)/\sqrt2$, 
$p_k^\lor=(p_k^x-ip_k^y)/\sqrt2$, 
and $p_k^+=(p^0_k+p^z_k)/\sqrt{2}$  
are momenta {\it entering} the diagram on leg $k$.
The coupling $g$ is proportional to the conventional QCD coupling $g_s$. 
Note that these are light-cone gauge ($A_-=0$) expressions
and include the contributions that arise when the longitudinal
field $A_+$ is eliminated from the formalism.
These rules are given in the context of  
't Hooft's $1/N_c$ expansion at fixed $N_cg^2_s$. 
Then the {\it planar} diagrams of the $SU(N_c)$ theory
are correctly given if we take $g\equiv g_s\sqrt{N_c/2}$. Non-planar diagrams
with this definition of $g$ must be accompanied by appropriate
powers of $1/N^2_c$, depending on the
number of ``handles'' in the diagram. Here we restrict attention
to planar diagrams, so our results should be compared
to the limit $N_c\to\infty$, fixed $g_s^2N_c$ of those in
the literature. In making such comparisons, note that 
our definition of $g$ multiplies conventionally
defined $n$-gluon tree amplitudes by a factor $N_c^{n/2-1}\to N_c$ for
$n=4$, so for each gluon scattering process we remove this factor
before comparing to the literature.

\section{Resolution of IR divergences using a discrete $p^+$ regulator}
It is well-understood that
a consistent resolution of infrared divergences in loop corrections 
to scattering amplitudes involves a cancelation in the 
rates against corresponding
infrared divergences in the rates for the emission (or
absorption) of an extra gluon, whose momentum is either collinear
with one of the gluons in the core process or ``soft''. 

\begin{figure}[ht]
\begin{center}
\psfrag{'1'}{$1$}
\psfrag{'2'}{$2$}
\psfrag{'3'}{$3$}
\psfrag{'4'}{$4$}
\includegraphics[width=4in]{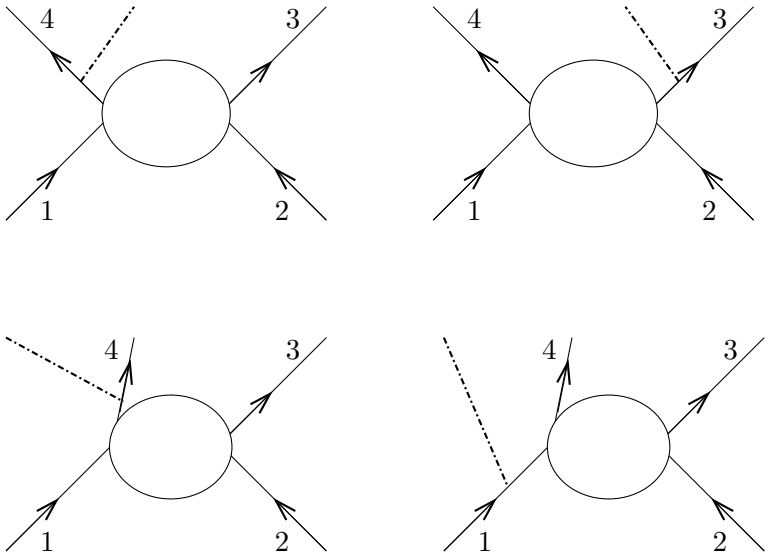}
\caption{The bremsstrahlung diagrams associated
with glue-glue scattering involving leg 4. At $N_c=\infty$
the sum of the diagrams on each line may be independently squared to give the
leading contribution to the cross section. Similar pairs of diagrams involving
each of the other legs must also be included.}
\label{brem}
\end{center}
\end{figure}

In the context of the large $N_c$ limit one need
combine coherently only Bremsstrahlung diagrams 
with the same cyclic ordering.
For example, in the diagrams shown in Fig.~\ref{brem} at $N_c=\infty$
it is only necessary to square the sum of the two diagrams on each line 
and combine the results on different lines incoherently.
Because $N_c=\infty$
suppresses nonplanar diagrams, it is convenient to
take an extra gluon line attached between two 
outgoing gluons
(as with the diagrams on the first line of Fig.~\ref{brem}) 
to be outgoing.
Similarly a gluon line attached between two incoming gluons 
is taken to be incoming. 
On the other hand, both outgoing and incoming extra gluons
must be considered when attached between an incoming and an outgoing
gluon (as with the diagrams on the second line of Fig.~\ref{brem}).

Infrared and collinear divergences occur
only when the Bremsstrahlung
gluon attaches to external legs. For example if the extra 
gluon is collinear
with $p_4$, there is a collinear divergence in the phase-space integral
of the square of the diagrams where the gluon is emitted from or absorbed by
leg 4. Calling the extra gluon's four-momentum $k$, for fixed $k^+$ the
collinear point is ${\bfs k}=k^+{\bfs p}_4/p_4^+$, and it is convenient
to write
\bea
{\bfs k}=k^+{{\bfs p}_4\over p_4^+}+{\hat{\bfs k}}
\eea 
and examine the phase space integral for $|{\hat{\bfs k}}|$ 
in a neighborhood of zero. Here we assume $k^+=O(1)$ so the
extra gluon is not soft. This is the customary jet interpretation
of the scattered gluon \cite{stermanw}. We define the jet resolution
$\Delta$ by the condition (\ref{jetdef}). 
This translates to ${\hat{\bfs k}}^2<|k^+|\Delta^2/|p_4^+|$.

The amplitudes for the emission of a hard 
collinear gluon from the right of leg 4 (as in the first diagram on
the first line of Fig.~\ref{brem}) are given, for the two polarizations, by
\bea
A_{\rm Brem}^\lor&=&-2g{k^++p_4^+\over k^+p_4^+}
{K^\lor_{k,4}A_{\rm Core}(p_1,p_2,p_3,k+p_4)\over
(k+p_4)^2},\qquad {\rm Outgoing~helicity}\\
A_{\rm Brem}^\land&=&
-2g{p_4^+\over k^+(k^++p_4^+)}{K^\land_{k,4}
A_{\rm Core}(p_1,p_2,p_3,k+p_4)\over
(k+p_4)^2},
\qquad {\rm Incoming~helicity}
\eea
When the extra gluon (with momentum $k$) is emitted from the left of leg 4, 
the amplitudes are the
same except that ${\bfs K}_{4,k}$ appears instead of ${\bfs K}_{k,4}$. Thus the
amplitudes for emission from left and right 
have opposite signs. The amplitudes do not cancel, however,
because they have different gauge group structure. At $N_c=\infty$ the two terms enter the cross section incoherently. When the
extra gluon has the same helicity as leg 4 and is collinear with $p_4$, 
it and gluon 4 are distinguished only by their $p^+$ values. Then we
arbitrarily call the one with smaller $|p^+|$ the extra gluon.

Now it is easy to see that
\bea
{\bfs K}_{k,4}&=&-p_4^+{\hat{\bfs k}},\qquad
(k+p_4)^2=-p_4^+{\hat{\bfs k}}^2/k^+=-2p_4^+{\hat{k}}^\land{\hat{k}}^\lor/k^+
\eea
Then we have, in $d$ transverse dimensions,
\bea
&&{d{\bfs p}_4\over 2|p_4^+|}{d{\bfs k}\over 2|k^+|(2\pi)^{d+1}}
(|A^\lor|^2+|A^\land|^2)
=\nonumber\\
&&\hskip+.5in{d{\bfs P}\over 2|P^+|}
{d{\hat{\bfs k}}\over|k^+| (2\pi)^{d+1}}\left({P^{+}-k^+
\over P^+}\right)^{d-1}\left({(P^+)^2
\over (P^+-k^+)^2}+{(P^+-k^+)^2
\over (P^+)^2}\right)
{g^2\over {\hat{\bfs k}}^2}|A_{\rm Core}|^2
\eea
where $P^\mu=k^\mu+p_4^\mu$.
The collinear divergence is now transparent in the integration over
${\hat{\bfs k}}$ near zero.  The coefficient of the
phase space factor ${d{\bfs P}/2|P^+|}$
combines nicely with the square of the
tree amplitudes with self-energy corrections on external
lines.
%%%%%%%%%%
The collinear divergence is present at finite $k^+$ and is not regulated
by our $k^+$ discretization. However, in lightcone gauge 
it cancels when combined with
the self energy correction on the corresponding external line.
To properly arrange this cancelation on-shell we need an
additional regulator.
In \cite{chakrabartiqt2} we introduced a temporary gluon mass. 
Here we regulate by sending $d\to2$ from above only after the
combination. For $d>2$, the required integral is simply 
\bea
\int_{0<{\hat{\bfs k}}^2|p_4^+|<|k^+|\Delta^2} {d{\hat{\bfs k}}\over
{\hat{\bfs k}}^2}&=&{1\over (d-2)}
{2\pi^{d/2}\over\Gamma(d/2)}\left({|k^+|\over|p^+_4|}\Delta^2\right)^{d/2-1}
\eea
Then the coefficient of the jet phase space factor is
\bea
\int_\Delta{d{\bfs k}\over 2|k^+|(2\pi)^{d+1}}(|A^\lor|^2+|A^\land|^2)
&&\\
&&\hskip-1.8in={g^2|A_{\rm Core}|^2\over|P^+|4\pi^2\Gamma(d/2)(d-2)}
\left({|k^+||P^+-k^+|\over|P^+|^2}\right)^{d/2-2}\left({\Delta^2\over4\pi}\right)^{d/2-1}
\left(1+{|P^{+}-k^+|^4
\over |P^+|^4}\right)
\eea
The blowup as $d\to2$ is the collinear divergence we are seeking
to resolve.
According to the Lee-Nauenberg theorem, to get an infrared safe
quantity we must sum over all $k^+$ in the range $0<|k^+|<|P^+|$.
And we must also include collinear emission from the
left of leg 4. The first term represents the emission of a gluon
with identical helicity to leg 4, so when we sum that term over the whole range
of $k^+$ we have included emission from both the left and right
of leg 4. However, the second term represents the emission of a gluon with
opposite helicity, and when summed over the whole range
gives only gluon emission from the right of leg 4. The emission
of an opposite helicity gluon (with momentum $k$) from the left has the
same squared amplitude, but it is convenient to switch the roles of
$k$ and $p_4$, so $k$ always refers to the right gluon.
Then the total emission rate is given by
\bea
\sum_{0<|k^+|<|P^+|}
\int_\Delta{d{\bfs k}\over 2|k^+|(2\pi)^{d+1}}(|A^\lor|^2+|A_R^\land|^2
+|A_L^\land|^2)
&=&\nonumber\\
&&\hskip-4in{g^2|A_{\rm Core}|^2\over|P^+|4\pi^2\Gamma(d/2)(d-2)}\sum_{k^+}
\left({|k^+||P^+-k^+|\over|P^+|^2}\right)^{d/2-2}
\left({\Delta^2\over4\pi}\right)^{d/2-1} 
\left(1+{|P^+-k^+|^{4}
\over |P^{+}|^4}+{|k^+|^{4}
\over |P^{+}|^4}\right)
\eea
Calling $x=|k^+|/|P^+|$, $1/x(1-x)$ times
the quantity in parentheses can be rearranged 
\bea
{1\over x(1-x)}+{(1-x)^3\over x}     
+{x^3\over1-x}&=&2\left(x(1-x)+{x\over 1-x}+{1-x\over x}\right)
\eea
So with this notation the squared amplitude for jet production
along gluon 4 is
\bea
\sum_{0<|k^+|<|P^+|}
\int_\Delta{d{\bfs k}\over 2|k^+|(2\pi)^3}(|A^\lor|^2+|A_R^\land|^2
+|A_L^\land|^2)
&&\nonumber\\
&&\hskip-3in ={g^2|A_{\rm Core}|^2\over|P^+|4\pi^2\Gamma(d/2)(d-2)}
\sum_{k^+} 
\left({1\over x(1-x)}+{x^3\over 1-x}+{(1-x)^3\over x}\right)
\left({x(1-x)\Delta^2\over4\pi}\right)^{d/2-1}\\
&&\hskip-3in ={g^2|A_{\rm Core}|^2\over|P^+|2\pi^2\Gamma(d/2)(d-2)}
\sum_{k^+} 
\left(x(1-x)+{x\over 1-x}+{1-x\over x}\right)
\left({x(1-x)\Delta^2\over4\pi}\right)^{d/2-1}
\label{collinearbrem}
\eea
In Ref.~\cite{chakrabartiqt2} 
we obtained an IR finite on-shell wave function renormalization
by introducing the same gluon mass used in the collinear emission
calculation. Instead, here we simply redo the self energy calculation
with $d>2$:
\bea
\Pi^{\land\lor}&=&-{g^2\over 4\pi^2}{p^2\over|P^+|}{1\over(4\pi)^{d/2-1}}
\sum_{k^+} 
\left(x(1-x)+{x\over 1-x}+{1-x\over x}\right)\int_0^\infty {dT
e^{-x(1-x)p^2T}\over (T+\delta)^{d/2}}\\
&\to&-{g^2\over 4\pi^2}{p^2\over|P^+|}{1\over(4\pi)^{d/2-1}}
\sum_{k^+} 
\left(x(1-x)+{x\over 1-x}+{1-x\over x}\right)\int_0^\infty {dT
\over (T+\delta)^{d/2}}\\
Z-1&\to&-{g^2\over 4\pi^2}{1\over|P^+|}
\sum_{k^+} 
\left(x(1-x)+{x\over 1-x}+{1-x\over x}\right){(4\pi\delta)^{1-d/2}
\over {d/2-1}}
\eea
Combining this wave function renormalization with the collinear
emission rate in $d>2$ transverse dimensions gives
%\vskip1in
\bea
\langle |{\cal M}|^2\rangle_{\rm jet}&&\nonumber\\
&&\hskip-.75in={g^2\over 4\pi^2}{|A_{\rm Core}|^2\over|p^+|}\sum_{k^+} 
\left(x(1-x)+{x\over 1-x}+{1-x\over x}\right)
{\left({x(1-x)\Delta^2/4\pi}\right)^{d/2-1}-\Gamma(d/2)
(4\pi\delta)^{1-d/2}\over(d/2-1)\Gamma(d/2)}\nonumber\\
&&\hskip-.75in\to{g^2\over 4\pi^2}{|A_{\rm Core}|^2\over|p^+|}\sum_{k^+} 
\left(x(1-x)+{x\over 1-x}+{1-x\over x}\right)
\ln\left({x(1-x)\Delta^2\delta e^\gamma}\right)
\eea 
for $d\to2$,
in complete agreement with the $\mu\neq0$ regulation method
of \cite{chakrabartiqt2}.

The collinear radiation discussed so far does not necessarily have
low momentum: only when $k^+$ or $P^+-k^+$ is small will this
radiation be soft. When it is soft, it is no longer valid to 
neglect the interference between diagrams where the soft gluon
is emitted from different external lines. In \cite{chakrabartiqt2}
my collaborators and I took this interference into account for
just that soft radiation that satisfied the collinear constraints
for at least one of the external lines. This was sufficient
to cancel all infrared divergences in the rates calculated
to one loop order. Moreover, since all the added radiation, 
both collinear and soft,
was constrained by Lorentz invariant constraints the final results
were Lorentz covariant. 

Here instead we include all soft Bremsstrahlung radiation
satisfying a single energy constraint in addition to the collinear
radiation. In the lightcone description it is natural to specify
the energy constraint in terms of the $+$ component of the 
extra gluon momentum: $k^+<\kappa$. We shall see that the rate
for this radiation is quite simple to calculate. To avoid 
double counting we must at the same time exclude these soft
gluons from the collinear calculation:
\bea
{\rm Soft~Radiation}:&&\qquad k^+<\kappa\nonumber\\
{\rm Collinear~Radiation}:&&\qquad \kappa<k^+<P^+-\kappa
\eea 
These prescriptions guarantee that there is no double counting.
One must bear in mind though that these constraints break
Lorentz invariance. 

In calculating the soft part of Bremsstrahlung, we must be sure
to combine coherently the two diagrams where the soft gluon
attaches to two neighboring lines in the same cyclic ordering.
For definiteness, take the coherent emission of a gluon
between legs $3$ and $4$, both of which we assume to have outgoing helicity.
Then the emission amplitudes are
\bea
A^\lor&=& -2gA_{\rm Core}\left[{k^++p_4^+\over k^+p_4^+}{K_{k,4}^\lor
\over (k+p_4)^2}+{k^++p_3^+\over k^+p_3^+}{K_{3,k}^\lor
\over (k+p_3)^2}\right]\\
&\sim&-{2gA_{\rm Core}\over k^+}\left[{K_{k,4}^\lor
\over (k+p_4)^2}-{K_{k,3}^\lor
\over (k+p_3)^2}\right]\\
A^\land&=& -2gA_{\rm Core}\left[{p_4^+\over k^+(k^++p_4^+)}{K_{k,4}^\land
\over (k+p_4)^2}+{p_3^+\over k^+(k^++p_3^+)}{K_{3,k}^\land
\over (k+p_3)^2}\right]\\
&\sim& -{2gA_{\rm Core}\over k^+}\left[{K_{k,4}^\land
\over (k+p_4)^2}-{K_{k,3}^\land
\over (k+p_3)^2}\right]
\eea
In these formulas we have assumed that $A_{\rm Core}$ is the same in 
both terms, which is approximately true since all components of
$k$ are small. 
The squared amplitudes for small $k^+$ are:
\bea
|A^\lor|^2\approx|A^\land|^2 &\sim&{2g^2|A_{\rm Core}|^2\over k^{+2}}
\bigg\{{{\bfs K}_{k,4}^2\over(k+p_4)^4}
+{{\bfs K}_{k,3}^2\over(k+p_3)^4}
-{2{\bfs K}_{k,3}\cdot{\bfs K}_{k,4}\over
(k+p_3)^2(k+p_4)^2}\bigg\}
\eea
We next use 
$$(k+p_4)^2=2{\bfs k}\cdot{\bfs p}_4-k^+{\bfs p}_4^2/p_4^+
-p_4^+{\bfs p}^2/k^+=-{\bfs K}^2_{k_4}/k^+p_4^+,\qquad
(k+p_3)^2=-{\bfs K}^2_{k_3}/k^+p_3^+$$
to write
\bea
|A^\lor|^2+|A^\land|^2 &\sim&{4g^2|A_{\rm Core}|^2}
\bigg\{{p_4^{+2}{\bfs K}_{k,3}^2+p_3^{+2}{\bfs K}_{k,4}^2
-2p_3^+p_4^+{\bfs K}_{k,3}\cdot{\bfs K}_{k,4}\over
{\bfs K}_{k,3}^2{\bfs K}_{k,4}^2}\bigg\}\nonumber\\
&\sim&{4g^2|A_{\rm Core}|^2}
{(p_4^+{\bfs K}_{k,3}-p_3^+{\bfs K}_{k,4})^2\over
{\bfs K}_{k,3}^2{\bfs K}_{k,4}^2}={4g^2|A_{\rm Core}|^2}
{k^{+2}{\bfs K}_{3,4}^2\over
{\bfs K}_{k,3}^2{\bfs K}_{k,4}^2}
\eea
The fact that this expression for soft gluon emission 
behaves as $1/{\bfs k}^4$ for large
transverse momentum is the reason that a soft constraint on 
$k^+$ imposes a soft constraint on all components of $k^\mu$.
Then the probability for emission of a soft gluon between legs 3 and 4 is
in $d$ transverse dimensions (Note that the integral over ${\bfs k}$.
 converges in both the IR and UV for $2<d<4$!)
\bea
|{\cal M}^{\rm Soft}_{34}|^2&=&{4g^2|A_{\rm Core}|^2}\sum_{k^+<\kappa}\int
{d^d{\bfs k}\over 2k^+(2\pi)^{d+1}}
{k^{+2}{\bfs K}_{3,4}^2\over p_3^{+2}p_4^{+2}({\bfs k}-k^+{\bfs v}_3)^2
({\bfs k}-k^+{\bfs v}_4)^2}\nonumber\\
&=&{4g^2|A_{\rm Core}|^2}\sum_{k^+<\kappa}
{k^{+}{\bfs K}_{3,4}^2\over 4\pi p_3^{+2}p_4^{+2}}
\int
{d^d{\bfs k}\over (2\pi)^d}
\int_0^1 dt{1\over[{\bfs k}^2+k^{+2}t(1-t)({\bfs v}_3-{\bfs v}_4)^2]^2}
\nonumber\\
&=&{4g^2|A_{\rm Core}|^2}\sum_{k^+<\kappa}k^{+d-3}
{{\bfs v}_{34}^2\over 4\pi}
\int_0^1 dt [t(1-t){\bfs v}_{34}^2]^{d/2-2}\int
{d^d{\bfs k}\over (2\pi)^d}
{1\over[{\bfs k}^2+1]^2}\nonumber\\
&=&{4g^2|A_{\rm Core}|^2}{[{\bfs v}_{34}^2]^{d/2-1}\over 4\pi}
\sum_{k^+<\kappa}k^{+d-3}
{\Gamma(d/2-1)^2\over\Gamma(d-2)}{\Gamma(2-d/2)\over(4\pi)^{d/2}}
\nonumber\\
&=&c_d{g^2|A_{\rm Core}|^2\over 4\pi^2}{[{\bfs v}_{34}^2]^{d/2-1}}
\sum_{k^+<\kappa}k^{+d-3}{2\over d/2-1},\qquad
c_d\equiv{\Gamma(d/2)^2\Gamma(2-d/2)\over\Gamma(d-1)(4\pi)^{d/2-1}}
\eea
where ${\bfs v}_k\equiv {\bfs p}_k/p^+_k$, ${\bfs v}_{kl}\equiv
{\bfs v}_k-{\bfs v}_l$, and  we have defined $c_d$, which goes
to 1 for $d=2$, for simplicity of writing. 
In this section keeping $k^+$ discrete serves as our
IR regulator. It is convenient to include the part of this soft radiation 
that also satisfies the collinear
constraint for one of the external legs in the collinear calculation,
so that the cancelation of the collinear divergence 
with the self-energy occurs for the full range of $k^+$.
The part of the leg 4 and leg 3 collinear emission contributing to
the 34 soft radiation is 
\bea
{g^2|A_{\rm Core}|^2\over2\pi^2\Gamma(d/2)(d-2)}\sum_{k^+<\kappa} 
{1\over k^+}
\left\{\left({k^+\Delta^2\over4\pi|p_4^+|}\right)^{d/2-1}
+\left({k^+\Delta^2\over4\pi|p_3^+|}\right)^{d/2-1}\right\}
\eea  
We check that for $d\sim2$
\bea
{1\over c_d\Gamma(d/2)(4\pi)^{d/2-1}}\sim {1+(d-2)\Gamma^\prime(1)
\over (1+3(d/2-1)\Gamma^\prime(1))(1-(d/2-1)\Gamma^\prime(1))}
=1+O((d-2)^2)
\eea
Subtracting the soft collinear rate from the soft rate we then find
%\vskip1in
\bea
|{\cal M}^{\rm Soft}_{34}|^2-|{\cal M}^{\rm Soft\&Col}_{34}|^2
&&\nonumber\\
&&\hskip-1.2in \sim c_d{g^2|A_{\rm Core}|^2\over 4\pi^2}
\sum_{k^+<\kappa}{1\over k^{+}}{1\over d/2-1}
\left(2[{k^{+2}{\bfs v}_{34}^2}]^{d/2-1}
-\left({k^+\Delta^2\over|p_4^+|}\right)^{d/2-1}
-\left({k^+\Delta^2\over|p_3^+|}\right)^{d/2-1}\right)\nonumber\\
&&\hskip-1.2in\to{g^2|A_{\rm Core}|^2\over 4\pi^2}
\sum_{k^+<\kappa}{1\over k^{+}}
\ln{k^{+2}{\bfs v}_{34}^4p_3^+p_4^+\over\Delta^4},\qquad d\to 2
\eea
As already mentioned, the Bremsstrahlung included in the
calculations of \cite{chakrabartiqt2} satisfied different constraints.
These were simply the union of the four regions $R_1\cup R_2\cup R_3\cup R_4$:
\bea
R_i:\quad {(p^+_i{\bfs k}-k^+{\bfs p}_i)^2\over |k^+P^+_i|}<\Delta^2 
\eea 
Avoiding double counting was a tedious headache, but eventually the
result for soft minus collinear radiation
assumed a reasonably compact form:
\bea
|{\cal M}^{\rm Soft,CQT}_{34}|^2-|{\cal M}^{\rm Soft\&Col,CQT}_{34}|^2
&=&
+{g^2|A_{\rm Core}|^2\over4\pi^2}
\sum_{|k^+|<\Delta^2/|P^+_4|v_{34}^2}
{1\over|k^+|}\ln{{k^{+2}{\bfs v}_{34}^4|p_3^+p_4^+|
\over \Delta^4}}\\
&&\hskip-1.5in\approx
+{g^2|A_{\rm Core}|^2\over4\pi^2}
\left[\sum_{|k^+|<A}
{1\over|k^+|}\ln{k^{+2}{\bfs v}_{34}^4|P_3^+P_4^+|
\over \Delta^4}-\ln{\Delta^2\over A|P_4^+|v_{34}^2}\
\ln{\Delta^2\over A|P_3^+|v_{34}^2}\right]
\label{34brem}
\eea
Here $A$ is chosen much larger than the $k^+$ discretization
unit. This formula is of course insensitive to the choice of $A$.
But by choosing $A=\kappa$ we find a very simple relation
between the Bremsstrahlung radiation calculated in the present
article and that calculated in \cite{chakrabartiqt2}.
\bea
|{\cal M}^{\rm Brem}_{34}|^2&=&|{\cal M}^{\rm Brem,~CQT}_{34}|^2
+{g^2|A_{\rm Core}|^2\over4\pi^2}
\ln{\Delta^2\over\kappa|P_4^+|v_{34}^2}\
\ln{\Delta^2\over \kappa|P_3^+|v_{34}^2}
\eea
Thus we can immediately write down the new probabilities for
glue glue scattering by making the appropriate adjustment
to the results of \cite{chakrabartiqt2}:
\bea
P^{CQT}_{\land\land\lor\lor}&=&|A_{\land\land\lor\lor}|^2
\bigg[1+\frac{g^2}{4\pi^2}\bigg[-2\log^2{\frac{\Delta^2}{s}}
-2\log^2{\frac{\Delta^2}{|t|}}-\frac{\pi^2}{3}+\frac{67}{9}\nn\\
&&\hskip2in-\frac{11}{3}\left[\log{(\Delta^2\delta e^{\gamma})}
+\log{\frac{\Delta^2}{|t|}}\right]+\log^2{\frac{s}{|t|}}\bigg]\bigg]\label{jetprobfinal}\\
&&\phantom{or}\nn\\
P^{CQT}_{\land\lor\land\lor}&=&|A_{\land\lor\land\lor}|^2
\bigg[1+\frac{g^2}{4\pi^2}\bigg[-2\log^2{\frac{\Delta^2}{s}}
-2\log^2{\frac{\Delta^2}{|t|}}-\frac{\pi^2}{3}
+\frac{67}{9}\nn\\
&&\hskip.5in-\frac{11}{3}\left[\log{(\Delta^2\delta e^{\gamma})}
+\frac{1}{2}\log{\frac{\Delta^4}{s|t|}}\right]
+\frac{(s^2+st+t^2)^2}{(t+s)^4}\log^2{\frac{s}{|t|}}\nn\\
&&\hskip1in
+\frac{(5st^2-5s^2t+11t^3-11s^3)}{6(t+s)^3}\cdot\log{\frac{s}{|t|}}
-\frac{ts}{(t+s)^2}\bigg]\bigg]
\label{jetprobfinal2}
\eea
We simply have to add the four terms
\bea
S&\equiv&\ln{\Delta^2\over\kappa|p_4^+|v_{34}^2}\
\ln{\Delta^2\over \kappa|p_3^+|v_{34}^2} 
+ \ln{\Delta^2\over\kappa|p_1^+|v_{12}^2}\
\ln{\Delta^2\over \kappa|p_2^+|v_{12}^2} \nonumber\\
&&+ \ln{\Delta^2\over\kappa|p_4^+|v_{14}^2}\
\ln{\Delta^2\over \kappa|p_1^+|v_{14}^2} 
+ \ln{\Delta^2\over\kappa|p_2^+|v_{23}^2}\
\ln{\Delta^2\over \kappa|p_3^+|v_{23}^2}
\eea
inside the square brackets. We use 
$$v_{ij}^2={(p_i+p_j)^2\over p_i^+p_j^+}={|(p_i+p_j)^2|\over |p_i^+p_j^+|}$$
and $|(p_1+p_2)^2|=|(p_3+p_4)^2|=|s|$, $|(p_1+p_4)^2|=|(p_2+p_3)^2|=|t|$,
to combine these terms with the first two terms in square brackets
\bea
S-2\ln^2{\Delta^2\over s}-2\ln^2{\Delta^2\over|t|}
&=&\ln{|p_3^+|\over\kappa }\
\ln{|p_4^+|\over \kappa } 
+ \ln{|p_2^+|\over\kappa}\
\ln{|p_1^+|\over \kappa} 
+\ln{\Delta^2\over s}\ \ln {\prod_i|p_i^+|\over\kappa^4}\nonumber\\
&&+\ln{|p_3^+|\over\kappa }\
\ln{|p_2^+|\over \kappa } 
+ \ln{|p_4^+|\over\kappa}\
\ln{|p_1^+|\over \kappa} 
+\ln{\Delta^2\over |t|}\ \ln {\prod_i|p_i^+|\over\kappa^4}\nonumber
\\
&=&\ln{|p_3^+p_1^+|\over\kappa^2 }\
\ln{|p_2^+p_4^+|\over \kappa^2 } 
-\ln{s|t|\over\Delta^4}\ \ln {\prod_i|p_i^+|\over\kappa^4}
\eea
Then our new results are
\bea
P_{\land\land\lor\lor}&=&|A_{\land\land\lor\lor}|^2
\bigg[1+\frac{g^2}{4\pi^2}\bigg[\ln{|p_3^+p_1^+|\over\kappa^2 }\
\ln{|p_2^+p_4^+|\over \kappa^2 } 
-\ln{s|t|\over\Delta^4}\ \ln {|p_1^+p_2^+p_3^+p_4^+|\over\kappa^4}\nonumber\\
&&\hskip2in-\frac{\pi^2}{3}+\frac{67}{9}-\frac{11}{3}\left[\log{(\Delta^2\delta e^{\gamma})}
+\log{\frac{\Delta^2}{|t|}}\right]+\log^2{\frac{s}{|t|}}\bigg]\bigg]\nn\\
&&\phantom{or}\label{jetprobnew1}\\
P_{\land\lor\land\lor}&=&|A_{\land\lor\land\lor}|^2
\bigg[1+\frac{g^2}{4\pi^2}\bigg[\ln{|p_3^+p_1^+|\over\kappa^2 }\
\ln{|p_2^+p_4^+|\over \kappa^2 } 
-\ln{s|t|\over\Delta^4}\ \ln {|p_1^+p_2^+p_3^+p_4^+|\over\kappa^4}\nn\\
&&\quad-\frac{\pi^2}{3}
+\frac{67}{9}-\frac{11}{3}[\log{(\Delta^2\delta e^{\gamma})}
+\frac{1}{2}\log{\frac{\Delta^4}{s|t|}}]+\frac{(s^2+st+t^2)^2}{(t+s)^4}\log^2{\frac{s}{|t|}}\nn\\
&&\qquad+\frac{(5st^2-5s^2t+11t^3-11s^3)}{6(t+s)^3}\cdot\log{\frac{s}{|t|}}
-\frac{ts}{(t+s)^2}\bigg]\bigg]
\label{jetprobnew2}
\eea
It is evident that this second definition of the  Bremsstrahlung
to be included in describing gluon scattering depends on the
Lorentz frame. This is in contrast to the first Lorentz invariant
definition. However it is more physically meaningful, because
it includes all radiation satisfying a single energy constraint
$k^+<\kappa$. This is particularly significant in high energy
scattering $s\to\infty$ at fixed $t$, the Regge limit. 
In the center of mass frame in the case that the scattering
plane is in the transverse direction, all the $|p_i^+|=\sqrt{s/8}$.
Then it is simple to see that the terms quadratic in $\ln s$
cancel. In contrast the first definition included so little
radiation at large $s$ and fixed $t$, 
that the coefficient of the $\ln^2 s$ term was negative.
\section{Comparison to Covariant Calculations.}
In order to make a definitive comparison of the noncovariant
lightcone gauge calculations of \cite{chakrabartiqt2}
to covariant calculations, it
is necessary to compare physical quantities. The elastic
amplitudes contain infrared divergences and calculations in
different gauges depend on the infrared cutoff:
they need not agree, and indeed they do not. 

We need to compare infrared safe quantities, such as the 
probabilities for jet scattering plus unobserved soft radiation.
The dimensionally regulated elastic one loop amplitudes
have long been available, e.g. in \cite{kunsztst}.
For comparison to our results we need to redo our
Bremsstrahlung calculations in dimensional regularization.
We have presented our intermediate results for general
transverse dimension $d>2$. We simply need to take $k^+$
continuous in the results and explicitly evaluate the $k^+$ integrals
at fixed $d>2$. The collinear jet production probability on
leg 4 becomes at continuous $\kappa<k^+<|p_4^+|-\kappa$
\bea
\sum_{\kappa<|k^+|<|P^+|-\kappa}
\int_\Delta{d{\bfs k}\over 2|k^+|(2\pi)^3}(|A^\lor|^2+|A_R^\land|^2
+|A_L^\land|^2)
&&\nonumber\\
&&\hskip-3.25in \to{g^2|A_{\rm Core}|^2\over2\pi^2\Gamma(d/2)(d-2)}
\int_{\kappa/|p_4^+|}^{1-\kappa/|p_4^+|} dx
\left(x(1-x)+{x\over 1-x}+{1-x\over x}\right)
\left({x(1-x)\Delta^2\over4\pi}\right)^{d/2-1}\nonumber\\
&&\hskip-3.25in \sim c_d{g^2|A_{\rm Core}|^2\over4\pi^2}
\int_{\kappa/|p_4^+|}^{1-\kappa/|p_4^+|} dx
\left(x(1-x)-2+{2\over x}\right)\left({1\over d/2-1}
+\ln x(1-x)\Delta^2\right)\nonumber\\
&&\hskip-3.25in \sim c_d{g^2|A_{\rm Core}|^2\over4\pi^2}
\left[\left({1\over d/2-1}
+\ln\Delta^2\right)\left(2\ln{|p_4^+|\over\kappa}-{11\over6}\right)
-\ln^2{|p_4^+|\over\kappa}-{\pi^2\over3}
+2\left(-{1\over4}+{1\over9}+2\right)\right]\nonumber\\
&&\hskip-3.25in \sim c_d{g^2|A_{\rm Core}|^2\over4\pi^2}
\left[\left({1\over d/2-1}
+\ln\Delta^2\right)\left(2\ln{|p_4^+|\over\kappa}-{11\over6}\right)
-\ln^2{|p_4^+|\over\kappa}-{\pi^2\over3}
+{67\over18}\right]\nonumber\\
&&\hskip1in+O\left({\kappa\over|p_4^+|},d-2\right)
\eea
There are of course similar expressions for the collinear
radiation from the other external legs.

Next we turn to the soft radiation between legs 3 and 4.
\bea
|{\cal M}^{\rm Soft}_{34}|^2&\to&
c_d{g^2|A_{\rm Core}|^2\over 4\pi^2}{[{\bfs v}_{34}^2]^{d/2-1}}
{2\over d/2-1}\int_0^{\kappa} dk^+k^{+d-3}=
c_d{g^2|A_{\rm Core}|^2\over 4\pi^2}{[\kappa^2{\bfs v}_{34}^2]^{d/2-1}}
{1\over (d/2-1)^2}\nonumber\\
|{\cal M}^{\rm Soft}_{34}|^2
&\sim&c_d{g^2|A_{\rm Core}|^2\over 4\pi^2}
\left[{1\over (d/2-1)^2}+{1\over (d/2-1)}\ln\kappa^2{\bfs v}_{34}^2
+{1\over2}\ln^2\kappa^2{\bfs v}_{34}^2\right]\nonumber\\
|{\cal M}^{\rm Soft}_{34}|^2
&\sim&c_d{g^2|A_{\rm Core}|^2\over 4\pi^2}
\left[{1\over (d/2-1)^2}+{1\over (d/2-1)}\ln{\kappa^2s\over|p_3^+p_4^+|}
+{1\over2}\ln^2{\kappa^2s\over|p_3^+p_4^+|}\right]
\eea
And there are three more such contributions associated with radiation
between legs 1,2; 1,4; and 2,3.

Collecting all the contributions to Bremsstrahlung radiation
gives
\bea
|{\cal M}^{\rm Brem,~total}|^2&\sim&c_d{g^2|A_{\rm Core}|^2\over4\pi^2}
\Bigg[\left({1\over d/2-1}
+\ln\Delta^2\right)\left(2\sum_i\ln{|p_i^+|\over\kappa}-{44\over6}\right)
\nonumber\\
&&-\sum_i\ln^2{|p_i^+|\over\kappa}-4{\pi^2\over3}
+4{67\over18}
+{4\over (d/2-1)^2}+{1\over (d/2-1)}
\ln{\kappa^8s^2|t|^2\over|p_1^+p_2^+p_3^+p_4^+|^2}\nonumber\\
&&
+{1\over2}\ln^2{\kappa^2s\over|p_1^+p_2^+|}
+{1\over2}\ln^2{\kappa^2s\over|p_3^+p_4^+|}
+{1\over2}\ln^2{\kappa^2|t|\over|p_1^+p_4^+|}
+{1\over2}\ln^2{\kappa^2|t|\over|p_2^+p_3^+|}\Bigg]\\
&\sim&c_d{g^2|A_{\rm Core}|^2\over4\pi^2}
\Bigg[{4\over (d/2-1)^2}+{1\over d/2-1}\left(\ln s^2|t|^2-{22\over3}\right)
\nonumber\\
&&+\left(\ln\Delta^2\right)\left(2\sum_i\ln{|p_i^+|\over\kappa}
-{22\over3}\right)
-\sum_i\ln^2{|p_i^+|\over\kappa}-4{\pi^2\over3}
+4{67\over18}\nonumber\\
&&
+{1\over2}\ln^2{\kappa^2s\over|p_1^+p_2^+|}
+{1\over2}\ln^2{\kappa^2s\over|p_3^+p_4^+|}
+{1\over2}\ln^2{\kappa^2|t|\over|p_1^+p_4^+|}
+{1\over2}\ln^2{\kappa^2|t|\over|p_2^+p_3^+|}\Bigg]
\eea
This must be added to the contribution of the one loop corrections
to the elastic scattering probability, which we take from \cite{kunsztst}
(We set their mass scale $\mu=1$)
\bea
P_{\wedge\wedge\vee\vee}^{\rm 1~Loop}&=&
c_d{g^2|A_{\rm Core}|^2\over4\pi^2}\bigg[|t|^{d/2-1}
\left(-{4\over(d/2-1)^2}+{1\over d/2-1}\left({11\over3}+2\ln{|t|\over s}
\right)+\pi^2-{67\over9}\right)\nonumber\\
&&\hskip4in+{11\over3}{1\over d/2-1}\bigg]\nonumber\\
&\sim&c_d{g^2|A_{\rm Core}|^2\over4\pi^2}\bigg[-{4\over(d/2-1)^2}
-{4\over d/2-1}\ln|t|-2\ln^2|t|+{1\over d/2-1}\left({11\over3}
+2\ln{|t|\over s}
\right)\nonumber\\
&&\hskip2in+\ln|t|\left({11\over3}
+2\ln{|t|\over s}\right) +\pi^2-{67\over9}+{11\over3}{1\over d/2-1}\bigg]
\nonumber\\
&\sim&c_d{g^2|A_{\rm Core}|^2\over4\pi^2}\bigg[-{4\over(d/2-1)^2}
+{1\over d/2-1}\left({22\over3}
-2\ln|t|s\right)\nonumber\\
&&\hskip2in+\ln|t|\left({11\over3}
-2\ln{s}\right) +\pi^2-{67\over9}\bigg]
\eea
Combining elastic plus Bremsstrahlung, the divergences as $d\to2$
cancel:
\bea
P_{\wedge\wedge\vee\vee}&\sim&c_d{g^2|A_{\rm Core}|^2\over4\pi^2}
\Bigg[\ln|t|\left({11\over3}
-2\ln{s}\right) -{\pi^2\over3}+{67\over9}
%\nonumber\\&&
+\left(\ln\Delta^2\right)\left(2\sum_i\ln{|p_i^+|\over\kappa}
-{22\over3}\right)\nonumber\\
&&
-\sum_i\ln^2{|p_i^+|\over\kappa}
+{1\over2}\ln^2{\kappa^2s\over|p_1^+p_2^+|}
+{1\over2}\ln^2{\kappa^2s\over|p_3^+p_4^+|}
+{1\over2}\ln^2{\kappa^2|t|\over|p_1^+p_4^+|}
+{1\over2}\ln^2{\kappa^2|t|\over|p_2^+p_3^+|}\Bigg]\\
&\sim&c_d{g^2|A_{\rm Core}|^2\over4\pi^2}
\Bigg[{11\over3}\ln{|t|\over\Delta^2} 
+\ln^2{s\over|t|}-{\pi^2\over3}+{67\over9}
%\nonumber\\&&
-{11\over3}\ln\Delta^2-\sum_i\ln^2{|p_i^+|\over\kappa}\nonumber\\
&&
+{1\over2}\ln^2{\kappa^2\over|p_1^+p_2^+|}
+{1\over2}\ln^2{\kappa^2\over|p_3^+p_4^+|}
+{1\over2}\ln^2{\kappa^2\over|p_1^+p_4^+|}
+{1\over2}\ln^2{\kappa^2\over|p_2^+p_3^+|}+\ln {s|t|\over\Delta^4}\
\ln{\kappa^4\over|p_1^+p_2^+p_3^+p_4^+|}\Bigg]\nonumber\\
&\sim&c_d{g^2|A_{\rm Core}|^2\over4\pi^2}
\Bigg[\ln^2{s\over|t|}-{\pi^2\over3}+{67\over9}
-{11\over3}\left(\ln\Delta^2-\ln{|t|\over\Delta^2}\right)
\nonumber\\&&\hskip2in
+\ln{\kappa^2\over|p_1^+p_3^+|}\ \ln{\kappa^2\over|p_2^+p_4^+|}
+\ln {s|t|\over\Delta^4}\
\ln{\kappa^4\over|p_1^+p_2^+p_3^+p_4^+|}\Bigg]
\eea
This result agrees in all respects with that obtained
from the discrete $k^+$ regularization (\ref{jetprobnew1}),
apart from the the dependence on the ultraviolet cutoff $\delta$,
which had been explicitly subtracted in the results presented
in \cite{kunsztst}.
It was already noted in \cite{chakrabartiqt2} that
the infrared insensitive ratio 
$P_{\wedge\wedge\vee\vee}/P_{\wedge\vee\wedge\vee}$
was in complete agreement with that presented in \cite{kunsztst}.
So we now have a definitive confirmation that the IR regulation 
supplied by the worldsheet lattice is completely equivalent to that
provided by dimensional regularization.

\section{Conclusion}
In this paper we have clarified two aspects of the glue-glue
scattering calculations of \cite{chakrabartiqt2}. 
First, we have
placed less restrictive constraints on the soft Bremsstrahlung radiation
which we combine with the one loop probabilities to cancel
infrared divergences. Since more soft radiation is allowed
by the new constraints, the scattering probabilities are increased 
over those obtained in 
\cite{chakrabartiqt2}. In particular, the large $s$ limit at
fixed $t$ now behaves as $\ln s$ instead of $-\ln^2 s$.
This behavior is compatible with Regge behavior.

Secondly, We extended our new calculations of Bremsstrahlung
to general continuous transverse dimensions. When 
combined with previous dimensionally regulated
covariant calculations of one loop corrections,
we obtained results in complete accord with those
regulated using the worldsheet lattice. This
is a much more detailed comparison than that made
in \cite{chakrabartiqt2}, which was limited by
the lack of a common treatment of Bremsstrahlung
radiation.
Thus, apart from collinear divergences, which are only occur for self
energy insertions on on-shell external lines, this discretization provides
a viable infrared regulator for lightcone calculations. 
\vskip14pt
\noindent\underline{ Acknowledgments}: 
This research was supported in part by the Department
of Energy under Grant No. DE-FG02-97ER-41029.

\end{document}